\documentstyle[11pt,aaspp4]{article} 
\newcommand{\kms}{km~s$^{-1}$}
\newcommand{\teff} {$T_{eff}$\/}
\newcommand{\subsun}{\mbox{$_{\odot}$}}
\newcommand{\etal}{{\it et al.\/}}
\newcommand{\eqw}{$W_{\lambda}$}
\newcommand{\cyg}{$\zeta$ Cyg}

\newcommand{\mc}[1]{\multicolumn{2}{c}{#1}}

\begin{document}

\title{An Abundance Analysis for Five Red Horizontal Branch Stars in the
Extremely Metal Rich Globular Cluster NGC 6553}

\author{Judith G. Cohen\altaffilmark{2},
        Raffaele G. Gratton\altaffilmark{3},
        Bradford B. Behr\altaffilmark{2} \&
        Eugenio Carretta \altaffilmark{3} }

\altaffiltext{1}{Based in large part on observations obtained at the
	W.M. Keck Observatory, which is operated jointly by the California 
	Institute of Technology and the University of California}
\altaffiltext{2}{Palomar Observatory, Mail Stop 105-24,
	California Institute of Technology, Pasadena, CA \, 91125,
        jlc@astro.caltech.edu, bbb@astro.caltech.edu}
\altaffiltext{3}{Osservatorio Astronomico di Padova, Vicolo dell'Osservatorio 5,
35122, Padova, Italy, gratton@pd.astro.it, carretta@pd.astro.it}

\begin{abstract}

We provide a high dispersion line-by-line abundance analysis
of five red HB stars in the extremely metal rich
galactic globular cluster NGC 6553. 
These red HB stars are significantly hotter than the
very cool stars near the tip of the giant branch in such
a metal rich globular cluster
and hence their spectra are
much more amenable to an abundance analysis than would be the case
for red giants.

We find that the mean [Fe/H] for NGC 6553 is $-0.16$ dex,
comparable to the mean abundance in the galactic bulge found
by McWilliam \& Rich (1994) and considerably higher than that
obtained from an analysis of two red giants in this cluster by
Barbuy \etal\ (1999).
The relative abundance for the best determined 
$\alpha$ process element (Ca) indicates an excess of $\alpha$ 
process elements of about a factor of two.
The metallicity of NGC 6553 reaches the average
of the Galactic bulge and of the solar neighborhood. 

\end{abstract}


\section{Introduction}

The abundances of the galactic globular clusters (henceforth GCs) are an important datum
for study of the chemical evolution and the halo of the Galaxy, among
many other areas.  
The determination of these abundances rests on spectroscopic observations
of the brightest globular cluster stars, the red giants at the tip
of the giant branch.
This subject has a very long history, with observations
of the brightest globular cluster stars attempted as soon as was
technically feasible, beginning with Helfer, Wallerstein \& 
Greenstein (1959) using the coud\'e of the Hale Telescope. 
In the late 1970s, the commissioning of a pair of identical echelle spectrographs 
attached to the newly 
completed $4-$m telescopes at KPNO and at CTIO
made it possible to obtain high signal-to-noise ratio (SNR) spectra at suitably
high spectral resolution for the
brightest giants in the nearer GCs, and several groups undertook
programs to provide the required abundance determinations (Cohen 1979, Pilachowski,
Sneden \& Wallerstein 1982, and references therein).
These programs produced excellent results for the more metal poor
GCs, but faltered at the metallicity of M71 or 47 Tuc for reasons described
in more detail below.  

As no very metal rich GC
is close enough for the giants to be reachable with instrumentation 
available at that time on a 4-m telescope, there was no
attempt to observe stars in any GC expected to be more 
metal rich than 47 Tuc. 
Cohen (1983) tried a few of the most metal rich GCs 
on the Palomar 5-m Hale telescope, but at somewhat lower spectral
resolution than ideal, to get preliminary abundance values.

Since there are roughly 150 galactic GCs, detailed abundance analyses
will only be available at best for a limited number of bright red giants 
in the nearer GCs.  Over the years, many schemes have been devised 
to extend the abundance determinations for individual red giants in
these few GCs to the full sample of galactic GCs.  Some approaches,
for example those of Zinn \& West (1984) and of Armandroff \& Zinn (1988),
use narrow band photometry or spectra of the integrated light
of GCs, some (see Armandroff \& Da Costa 1991) use the very strongest
lines in individual GC red giants measured from low dispersion spectra,
while other abundance schemes rely on 
characteristics of the c-m diagram, such as the 
color of the giant branch (Frogel, Cohen \& Persson 1983) or the
slope of the
giant branch in the infrared (Kuchinski \& Frogel 1995).
But the fundamental calibration of all these schemes rests on the high dispersion
spectroscopic abundance determinations for a limited number of individual
GC red giants.  Independent of these is the $\Delta(S)$ method for
RR Lyrae stars developed by Preston (1959), with a recent calibration
from high dispersion abundance analyses of field RR Lyrae stars by
Clementini \etal\ (1995).  Unfortunately, there are no RR Lyrae stars in the
most metal rich GCs, hence this is only useful for the metal poor GCs.

Up to the present the calibration of all these schemes at the 
extremely metal rich end has been unsatisfactory, a situation we hope to remedy
with the present work.
NGC 6553 is the best GC to use as a calibrator for the very
metal rich GCs.  It is reasonably close, populous, very metal rich with an 
abundance that is believed to lie near the maximum achieved among the galactic GCs, 
and with a reddening which is small compared to that of most such clusters.
The purpose of this paper is to present a detailed abundance analysis
for stars, avoiding the red giants, in NGC 6553.

\section {The Characteristics of NGC 6553}

The first photometric study of NGC 6553 was that of Hartwick (1975).
Ortolani, Barbuy \& Bica (1990) provide a modern
ground based CCD study.  Their key
finding was that the giant branch in NGC 6553 as viewed in a 
color-magnitude diagram using a blue magnitude such as
$V, B-I$ or $V, V-I$, is not monotonic, but forms
an arc such that the reddest of the presumed 
red giants in NGC 6553 are fainter than the somewhat bluer red giants.
This anomalous behavior is also seen in their later study of 
NGC 6528 (Ortolani, Barbuy \& Bica 1991) and in the galactic bulge field
giants (Rich \etal\ 1998).  It
is less prominent in c-m diagrams involving 
redder colors, i.e. $I, V-I$.   It is not
seen at all in c-m diagrams involving only
infrared wavelengths  where the giant branch behaves as expected,
with the stars becoming brighter as they become redder
(Cohen \& Sleeper 1995, Kuchinski \& Frogel 1995).
This anomalous behavior is ascribed to the extreme
line/molecular band absorption found at blue wavelengths in such
metal rich cool stars.

Guarnieri \etal\ (1998)
provide a definitive photometric study of NGC 6553 which combines 
$V$ and $I$ photometry from HST with
ground based infrared photometry at $J$ and $K$.  They 
determine a distance based on the
magnitude of the horizontal branch (henceforth HB) stars 
for this GC of 5.2 kpc with a reddening of $E(B-V)$ = 0.7 mag,
in good agreement with that determined by Zinn (1980) ($E(B-V)$ = 0.78 mag). 
Guarnieri \etal\ (1998) obtain the reddening 
by forcing the red giant branch of NGC 6553 to
overlay that of 47 Tuc, and then adopting a correction for the
difference in metallicity between the two GCs; to deduce this
correction they adopt [Fe/H] = $-0.22 {\pm}0.05$ dex for NGC 6553.
\footnote{We adopt the usual spectroscopic notation that [A/B] $\equiv$
log($N_A/N_B)_{star} - {\rm log}(N_A/N_B)$\subsun.  Also, unless otherwise
specified, metallicity is arbitrarily defined as the stellar [Fe/H] value.} 

Barbuy \etal\ (1992)
compiled the abundance estimates for NGC 6553 over the period
1983 through 1992 from the literature.
The values range from +0.5 to $-0.7$ dex, with none of them being
particularly well calibrated or having adequate spectral dispersion
to produce a highly accurate result.  Barbuy \etal\ (1992)
obtained a CCD echelle spectrum from ESO covering the spectral region
475 -- 580 nm with
a spectral resolution of 0.20\AA\ and a SNR of 40 of a single red
giant in NGC 6553 (star III--17 as identified by Hartwick 1975), which is
among the brightest red giants in the cluster (at $V$).  They
attempted an abundance analysis for this star.
The available photometry for this star could not be used
to determine \teff\ as different colors
gave inconsistent results.  The line crowding in their
spectrum was so severe that
they were forced to use spectral synthesis techniques throughout.  
They obtained a preliminary result of [M/H] = $-0.2$ (+0.2,$-0.4$) dex for 
this red giant in NGC 6553.

A detailed abundance analysis of two giants in NGC 6553 
(stars II--85 and III--3,
both slightly fainter than star III--17, but still among the brightest stars
in the cluster at $V$)
was carried out by Barbuy \etal\ (1999).  This time the spectral
region was shifted somewhat to the red, 500 -- 750 nm, and
was covered at a spectral resolution of R = 20,000 with a SNR $\approx$50.  The
effective temperatures for the two red giant stars were derived 
from the photometry of Guarnieri \etal\ (1998) and are 4000 K for both stars with
log(g) = 0.8.  They obtained [Fe/H] = $-0.55 {\pm}0.2$ dex with
[$\alpha$/Fe] $\approx +0.5$ dex.

\section{The Choice of Stars for Our Sample}

\subsection{Lessons From the Past}

In the early 1980s, the initial surge of enthusiasm for abundance determination
of galactic GCs with the new generation of telescopes and instrumentation
led to detailed results that have withstood the test of time for
metal poor GCs.  But when efforts were made to extend this upward in metallicity
to 47 Tuc and to M71, very controversial abundances were initially derived 
by both of the two groups involved
that were considerably lower than anyone anticipated ($-$1.1 dex for 47 Tuc 
by Pilachowski, Canterna \& Wallerstein 1980 and 
Pilachowski, Sneden \& Wallerstein 1983, and 
$-$1.3 dex for M71 by Cohen 1980).
Subsequently Cohen (1983) attempted to determine whether these low
abundances for 47 Tuc and for M71 were real or resulted
from some problem in the analysis.  Cohen (1983) derived [Fe/H] = $-$0.7 dex 
for M71 using (for perhaps the first time for a GC star) a CCD spectrum 
and demonstrated 
that the abundances initially deduced for stars in M71 (and 
presumably 47 Tuc as well) 
were too low.  This was ascribed to the difficulty of locating the proper
continuum level in such cool high metallicity stars and the tendency
towards underestimating the continuum in such heavily blended spectra. 
Gratton, Quarta \& Ortolani (1986) also obtained [Fe/H] = $-0.8$ for M71
and for 47 Tuc, observing stars considerably warmer (and fainter) than
the tip of the red giant branch, demonstrating yet again that 
there were problems with the
initial abundances for these clusters obtained 
from the photographic 4$-$m echelle spectra.
Note that the red giants near the tip of the giant branch in M71 and in 47 Tuc 
have \teff $\sim 4100$ K with log(g) ${\approx}0.7$ dex, somewhat
hotter than the two red giants
in NGC 6553 analyzed by Barbuy \etal\ (1999).

After removing the reddening, the giant branch of NGC 6553 is even redder than
that of 47 Tuc or M71 in a color-magnitude diagram.  Hence presumably
the giants near the tip are even cooler and 
more metal rich than those of M71.  Issues of line crowding and
continuum determination must be taken very seriously.  We must take 
every precaution to avoid stumbling over the same problem again.

\subsection{Our Choice of Stars in NGC 6553}

We are fortunate to have available the collecting power of the Keck Telescope
and the efficient HIRES spectrograph (Vogt \etal\ 1994).  This
means that we can get reasonable spectral resolution and SNR 
in a one hour exposure for
stars considerably fainter than the tip of the RGB in NGC 6553.
Instead of using red giants, even giants substantially fainter 
than the tip of the giant branch,
we have chosen stars on the red HB in NGC 6553 for our sample.
There are a number of major advantages of such a choice.  The first is
that the red HB stars are considerably hotter than those near the top
of the RGB, with \teff\ comparable to the luminous giants in more metal
poor globular clusters.  We can use the same set of absorption features
as were used for the lower metallicity GCs and they will be of roughly
comparable strength in a red HB star in NGC 6553 as in a RGB star in
a lower metallicity GC.  We will thus be more able to maintain
consistency of results over the full range of metallicity of
the galactic GCs.  

Moving into the near IR (630 - 870 nm with incomplete 
coverage due to gaps between the orders of HIRES) at high resolution
(R = 34,000, corresponding to a 1.1 arcsc entrance slit)
guarantees the minimum line crowding possible, while still avoiding
excessive contamination by terrestrial atmospheric absorption and emission lines.
We will see later that the resulting red HB star HIRES spectra, even in a 
GC as metal rich as NGC 6553, are sufficiently clean that we do not
need spectral synthesis techniques and can instead determine abundances
using individual absorption features as is normally done for the RG stars
in the metal poor GCs.
Furthermore with the powerful equipment of the Keck Telescope + HIRES
we can achieve a SNR of ${\approx}70$ per 4 pixel resolution
element with exposure times of under 1 hour
for a HB star in NGC 6553 in that spectral region.
 
Other advantages of working with red HB stars include a well defined
mass (assumed to be 0.8 M\subsun) and luminosity so that
once \teff\ is specified, the surface gravity is determined as well.
Furthermore, the use of a well populated evolutionary state that occupies
only a small area on a cluster c-m diagram
such as the red HB maximizes the probability that a star in that location
in the c-m diagram is a member of
the GC and minimizes contamination by field stars, a non-trivial
issue for a cluster with ($l^{II}, b^{II}$) = (5.3, $-3.0$).  In addition,
the non-LTE corrections are smaller for these hotter and higher surface gravity stars.

With these points in mind, we chose the least crowded red HB
candidates in NGC 6553 from Guarnieri \etal\ (1998) to form the
spectroscopic sample.  

At the request of A. Renzini, two of the extremely red stars at the extreme
end of the arc in the cm diagram for NGC 6553 found by Ortolani, Barbuy \& Bica (1990)
were also observed in order to check that they are indeed members of the cluster.
Since the spectra of these stars show very strong TiO bands, they were not
used for the abundance analysis.

Table 1 lists the five red HB stars in NGC 6553 observed spectroscopically
during a night in the summer of 1995 and one in the summer of 1997, as well
as the two extremely red giants.  The identification numbers are those
assigned by Guarnieri \etal\ (1998), from which the photometry is taken.
Star 30297, one of the extremely red stars,
is star II--3 of Hartwick (1975).   The exposure times and dates of observation
are also given.

The heliocentric radial velocity for this
cluster was not well determined, hence measurements from
our spectra are also given in Table 1.  We find that all of the stars
observed, including the two extremely red stars, are
probably members of NGC 6553.  The seven heliocentric radial velocities indicate
that the cluster mean velocity is 4 \kms, with $\sigma = 7.1$ \kms, in good agreement
with the recent determinations of the mean cluster velocity 
from two stars of Barbuy \etal\ (1999) of 6 \kms\
and (from more stars, but at lower dispersion) of 
Rutledge \etal\ (1997a) of 8.4 \kms.  The
HB stars are all slow rotators; this will be discussed in B. Behr's thesis.

There is one possible disadvantage of using horizontal branch stars.  Since
they are in an even later stage of evolution than RGB stars, one might worry
about the effects of mixing, gravitational settling, etc.
In the stable atmosphere of 
a hot star, diffusion effects -- both gravitational settling of He, and 
radiative levitation of Fe and other metals -- can significantly alter the 
photospheric composition, as predicted by Michaud, Vauclair \& Vauclair
(1983), and observed by Heber (1987) and Glaspey \etal\ (1989). These effects,
however, appear to be limited to \teff $\ge 10000$ K 
(Behr \etal\ 1999) and are thus 
not of concern for the RHB. Deep mixing on the RGB can also produce surface 
abundance anomalies in CNO (Kraft \etal\ 1998 and many prior reports), but 
these same variations will appear in any evolutionary stage
beginning with the RGB and including any more advanced stage such as the HB.
Such abundance anomalies are not expected to extend to the heavier elements
from Ca to the Fe peak.

There have been a small number of high dispersion
abundance analyses of RR Lyrae stars (Clementini \etal\ 1995 for M4)
and other HB stars in galactic GCs (Cohen \& McCarthy 1998 for M92)
which demonstrate that, at least if one avoids the bluest HB
stars, good consistency
with abundances from giant branch stars in the same GC is obtained.

\section{Determination of Atmospheric Parameters}

The initial values of \teff\ to be used in the high dispersion
abundance analysis are determined for each of the red HB
stars from the observed $V-K$
colors of Guarnieri \etal\ (1998) using the latest version of the
of model stellar atmospheres from Kurucz (1992) and the latest update of the
original calibration of Cohen, Frogel \& Persson (1978) which is given
by Gratton, Carretta \& Castelli (1997).
As has been discussed extensively in the past, the $V-K$ color offers
maximum leverage in assigning \teff.
A reddening of $E(B-V) = 0.78$ mag corresponding to
$E(V-K)$ = 2.12 mag is assumed; we use the reddening curve of Cohen \etal\ (1981). 
We further assume
that the mass of a HB star in NGC 6553 is 0.8 M\subsun\ and that the
luminosity of a star on the red horizontal branch in NGC 6553 is
log(L/L\subsun) = 1.62.  These then determine the surface gravity.
The resulting atmospheric parameters are listed in Table 2.  

\section{Equivalent Width Measurements}

The spectra were reduced using the suite of routines for analyzing 
echelle spectra written by McCarthy (1988) within the Figaro 
image processing package
(Shortridge 1988).  The stellar data are flat fielded with quartz lamp spectra, thereby
removing most of the blaze profile, and the results are normalized
to unity by fitting a 10th-order polynomial to line-free regions of
the spectrum in each order.

We are trying to measure equivalent widths for weak lines from 
spectra which are of adequate resolution ($\lambda/\Delta(\lambda)$ = 34,000)
but of only marginally adequate SNR (70/4 pixel resolution element).  Even
in the near infrared there will be some adjacent weak lines
contaminating any attempt to measure equivalent widths.  We therefore
adopt the following procedure to make the \eqw\
more stable to the presence of weak spectral features near the
desired one.  We smooth the spectra with a Gaussian whose FWHM is 0.3 \AA,
which is slightly larger than the instrumental resolution near 8000 \AA.  The
definition of the continuum was then updated slightly as seemed appropriate.  
Equivalent widths were
measured in the convolved spectra of HB stars 40071 and 40123 in NGC 6553
for about 40 lines spanning the full range in strength  to be used
and judged to be unblended.  A linear relationship
was established between \eqw\ and central depth 
of an absorption line, and this relationship
was used to determine \eqw\ for the rest of the features used
in the abundance analysis.  This process was repeated for
the three HB stars in NGC 6553 which were observed in 1995.
The fit between \eqw\ and FWHM determined for the spectra from 1995
is very slightly different from that of the 1997 data, as might be expect 
due to differing instrumental and/or telescope focus.  
In total, \eqw\ measurements were obtained for $\sim$80 lines 
in each of the five HB stars in NGC 6553, including $\sim$40 Fe I lines.
The \eqw\ we obtained for the five RHB stars in NGC 6553 are listed in Table 3.

Figure 1 shows a montage of a section of the convolved spectra covering
part of a single echelle order for the
five NGC 6553 RHB stars and for also the convolved spectrum
for the comparison star \cyg.  Our use of convolved spectra 
to measure \eqw\ and the fact that the 
stars involved have the low surface gravities characteristic of giants
means that any small variations in the intrinsic line profile
due, for example, to variations in thermal width or to a small amount of
hyperfine structure, will not significantly affect our measurements of \eqw.

Several checks have been made to verify that our procedure produces
accurate measurements of \eqw.  All of the five red 
HB stars in NGC 6553 have have very similar
atmospheric parameters (see Table 2).  We compare the measured \eqw\
for the pair of stars 30180 and 30242
and find $W_{\lambda}(30242) = 0.96\times W_{\lambda}(30180) -3.0$ m\AA\,
with a rms scatter about this relationship of 14.2 m\AA.  Attributing equal
errors to both sets of \eqw\ suggests that typical errors in
\eqw\ are $\pm$10 m\AA\ which arise largely from uncertainties
in the location of the continuum.

A second check can be made using spectra of much brighter comparison stars
taken with the identical configuration of HIRES on the same night as 
the NGC 6553 spectra.  These were chosen to have well known
abundances and at least some of
them are believed to be similar to the RHB stars in NGC 6553.  The one we utilize
in this paper is \cyg\ (BS 8115).  The spectrum of this star (which is also
included in Figure 1) has the
same resolution as the RHB stars in NGC 6553, but much higher SNR.  Thus the
set of \eqw\ could be measured on both the original spectrum and the convolved
spectrum; the \eqw\ for the latter are given in the last column of Table 3 and 
in Figure 2 these two sets of \eqw\ are compared.
The regression line between \eqw\ measured in the original and convolved 
spectra is (for 47 lines)
\eqw(conv) = $(1.033 \pm0.026) \times$ \eqw(orig) $- (2.1 \pm8.1)$ m\AA, 
where the error on the constant term is the rms scatter of individual 
points along this regression line. If we consider the set of \eqw\ measured on
the original spectra as ``perfect'', there is no clear systematic error
in the set of \eqw\ measured in the convolved spectra, and typical errors
are $\pm$8 m\AA. This estimate agrees quite well with the error of 
$\pm$10 m\AA\ estimated for the stars in NGC 6553 (where the SNR is lower).

\section {Abundance Analysis}

The guiding principle we have followed in the abundance analysis of the
RHB stars in NGC 6553 is to adhere as closely as possible to the procedures
used in earlier analyses of the less metal rich GCs by Carretta \& Gratton (1997)
(henceforth CG)
so as to make the results for the NGC 6553 sample as consistent and comparable
as possible with the earlier analyses.  This means we use the same set of
model atmospheres (those of Kurucz 1992 with convective overshooting), 
the same relationship between
color and \teff\ (that of Gratton, Carretta \& Castelli 1997), the same set of
gf values whose sources are given in CG, 
the same line-by-line abundance code, etc.  
The solar abundances are those
defined in CG, and are obtained in a manner as consistently as possible as are
those of the program stars, using a solar model from Kurucz (1992), etc.

The metal abundance for the model atmosphere used for each RHB star
was determined by looking at the results
of a first iteration for [Fe/H], then repeating the analysis with models 
whose abundance is close to that inferred from the first iteration.
The microturbulent velocity was determined in the usual manner
by forcing the slope of the
\eqw\ versus deduced abundance to be zero.  No hyperfine structure corrections
were used; none are necessary except perhaps for Sc II.  

Clementini \etal\ (1995) demonstrate that the non-LTE corrections for
HB stars are expected to be small.  We assume LTE, except for oxygen.
The O abundance is from the infrared triplet near 7770 \AA;  a small 
non-LTE correction ($-0.07$ dex) has been applied for oxygen based 
on the calculations of Gratton \etal\ (1999).

We now consider the validity of our initial atmospheric parameters as judged by
the results of the abundance analysis.  For three of the five RHB stars in NGC
6553, adoption of the initial parameters led no detectable trend of abundance versus
excitation potential for the Fe I lines.  For stars 40123 and 30257, 
this was not true;  to achieve this condition,
a small adjustment was made in their \teff, raising it by 150 to 200K.
This change could easily arise if the reddening were slightly 
larger than the mean for these two RHB stars.  An increase in
$E(B-V)$ by 0.08 mag above our adopted mean value for NGC 6553 of
0.78 mag is required for star 40123, while a somewhat smaller increase
is needed for star 30257.
With these two small adjustments of \teff,
the derived abundances are independent of the
excitation potentials of the FeI lines and the ionization equilibrium for Fe is
quite good (see Table 4).

The resulting abundances deduced for the five RHB stars in NGC 6553 are listed
in Table 4.  First the deduced [Fe/H] value is given as determined from the ions
Fe I and Fe II.  Then the element ratios [A/Fe] are given as determined from various
ions.  These are calculated using Fe I, except for oxygen and singly ionized scandium, 
where Fe II is used. 
The final column gives the solar abundance we use.  When there is more than one
line of a given ion, the rms dispersion about the mean is given in parentheses.
The mean abundance of each element together with the rms dispersion about the
mean for the cluster NGC 6553 as determined from the analysis of the
five red HB stars is given in Table 5.

Figure 3 shows the deduced abundances for Fe I lines as a function of 
excitation potential in the five red HB stars in NGC 6553.  The 
1$\sigma$ rms uncertainty
in the slope of the best linear fit is 0.026 dex/eV, implying an uncertainty
in \teff\ of $\sim$100 K.

\subsection{Verification using \cyg}

We observed several bright stars as possible comparison stars for this analysis
of RHB stars in the extremely metal rich GC NGC 6553.  Of those, \cyg\ (BS 8115)
is the one actually used as its atmospheric parameters are closest 
to those of  the program stars.  Just as for the GC stars, we 
obtain \teff\ for \cyg\ from its colors, including the $V-K$ color,
given by Johnson (1964) and by Neugebauer \& Leighton (1969)
using the same calibration
as for the RHB stars.   This star is so nearby
that interstellar reddening can be ignored.
The Hipparcos parallax (ESA 1997) is 21.62 mas,
implying $M_V = -0.13$ mag, somewhat brighter than the HB for old
stellar systems.  However, we do not know the mass of \cyg, 
and hence must determine
its surface gravity by forcing ionization equilibrium for 
Fe using the set of \eqw\ measured
from the original non-convolved spectra, and checking that the implied
mass is reasonable.
The abundance for the model
atmosphere was set in a manner similar to that used for the RHB stars, and an
abundance analysis was carried out both for the set of \eqw\ measured from the
original spectra and for those from the convolved spectra. The resulting 
abundances are listed in Table 6.  The organization of Table 6 is similar
to that of Table 4 described in detail above.

\cyg\ has been the subject of at least five relatively recent high dispersion
abundance analyses according to the compilation of 
Cayrel de Strobel \etal\ (1997).  The
\teff\ used in these five abundance analyses range from 4890 to 4990 K 
and log(g) ranges from 2.0 to 2.9.  The
deduced [Fe/H] ranges from $-$0.17 to +0.10 dex, with typical errors quoted as
$\pm$0.10 dex.  The mean of the five abundance determinations for \cyg\ 
from the literature is $-$0.04 dex.  Our adopted \teff\ is in the middle  
of the range of these
values, as is our log(g);
our value of [Fe/H] from Fe I is +0.05 dex.
This is in some sense the expected
abundance for a disk star.  Thus we feel that this
comparison demonstrates the validity of our procedure for measuring \eqw\
and of our abundance analysis procedure.

\subsection{Verification Using Spectral Synthesis}

To demonstrate how well the observed spectra can be fit by spectral
synthesis with the model atmosphere parameters we have derived
and with the abundances obtained from our analysis, we display in Figure 4 
a synthesis for the spectral region shown in Figure 1 for the
five RHB stars in NGC 6553.   In considering the different abundances
shown ($-0.4, -0.2$ and 0.0 dex) all metals (except Li)
were scaled appropriately
and a Ca overabundance of a factor of two ([Ca/Fe] = +0.3 dex) was
used throughout.  Figure 5 shows the same for the comparison star
\cyg, except that here the spectral syntheses shown utilize the
abundances $-0.2, 0.0$ and +0.2 dex.  The convolved spectra are 
shown for the NGC
6553 stars while the original spectrum is shown for \cyg.

The comparison with the predictions of spectral synthesis roughly confirm 
an  abundance of
about [Fe/H] = $-0.2$ for NGC 6553 
([Fe/H] = $-0.4$ appears definitely too low, while a solar
Fe abundance is clearly too high), in agreement with the abundance
given in Table 4 from the line-by-line analysis.  For \cyg, a
higher abundance is indicated, as we expect from the discussion above.

\subsection{Discussion of Errors}

The dominant source of errors is in the \eqw\ because of the 
limited SNR of our spectra of such relatively faint metal rich cool stars, 
which gives rise to an uncertainty
in the continuum location.  Also contributing are problems of cosmic
ray hits and accurate subtraction of night sky emission features, which may
perturb individual lines.  Another important factor is the uncertainty in
\teff\ due to possible reddening variations across NGC 6553.  
As mentioned earlier, a change in $E(B-V)$ from 0.78 mag to 0.86 mag
will produce a increase in \teff\ of 200 K.

Figure 1 of Hartwick (1975) demonstrates that the outer isophotes of NGC
6553 (i.e. the stellar distribution) are highly non-circular,
presumably the result of non-uniform interstellar absorption.
The NE quadrant in particular shows evidence for absorption
that is higher than the mean.  Cohen \& Sleeper (1995) present
evidence for a significant dispersion of
the interstellar absorption in several other highly reddened GCs 
by analyzing the dispersion in color of
the red giant branch measured at a fixed luminosity
for a variety of colors.  While NGC 6553 is not in their sample, 
NGC 6528, with a comparable $E(B-V)$, is included.  The full width in
$V-K$ of the upper giant branch in this cluster is $\sim$0.6 mag.
Ascribing this to reddening variations across the cluster implies
a range in $E(B-V)$ of $\sim$0.2 mag.  Ortolani, Barbuy \& Bica (1991)
and Ortolani, Bica \& Barbuy (1991) also
noted reddening variations
across the face of this GC of $\Delta(E(B-V) \sim 0.3$ mag.
Since these are full amplitudes, not rms dispersions, for the variation
in reddening, the fact that two stars in NGC 6553 appear to have a somewhat
larger reddening than the mean by about $\Delta[E(B-V)] \sim 0.08$ mag should
not be surprising.  However, it does indicate that considerably larger
variations in $E(B-V)$ should not be seen within our relatively small sample.

For that change in \teff\ of +200K, without changing any
other parameters, the Fe ionization equilibrium changes
by $-0.33$ dex, and [Fe/H] (from the more numerous FeI lines) increases
by +0.10 dex.  [O/Fe I] is quite sensitive to \teff, but when it is compared
with the dominant ionization stage Fe II, which
unfortunately has far fewer measurable lines in this spectral region 
for these stars than does Fe I, the O abundance is relatively
insensitive to a 200 K change in \teff.  However, in practice, part of any increase in 
\teff\ will be taken up by adjustments of other parameters, 
particularly the microturbulent velocity.
This is because there is a (loose) correlation between
line strength and excitation in our list of measured spectral lines,
with the lowest excitation lines in general being stronger.
For this reason the impact of
a change of \teff\ on the derived abundances is much less than the value
quoted above. We verified this by repeating our analysis with \teff\ lower
by 200 K, 
and adjusting $v_t$ again to 
avoid trends of derived abundances with \eqw. The average abundance we found 
for NGC 6553 was again [Fe/H] = $-0.16$.
Furthermore, the ionization equilibria we have
obtained just with our initial choice of model atmosphere parameters
for three of the five red HB stars in NGC 6553 are extremely good.

Model atmospheres whose detailed abundances element-by-element are identical to the
results we have found for NGC 6553 are not available.
We have checked that the error introduced by using models with the solar
abundance ratios is small, less than 0.1 dex for all ions.

In summary, we feel that
an uncertainty of ${\pm}0.1$ dex is reasonable for [Fe/H] and ${\pm}0.15$ 
dex for the element-to-element abundance ratios in each of the five
RHB stars in NGC 6553, with the caveat that the
O abundance be compared to that of the dominant singly-ionized
ionization stage of Fe.

\subsection{Element Ratios}

In our discussion to this point we have concentrated on
Fe I as it is the only ion for which our red spectra contain large numbers of 
measured lines.  For all other ions the number of measured lines is small.
Typical errors in abundances from individual lines are $\pm0.2$ dex, 
and are dominated by uncertainties in the location of the local continuum and by the
presence of blends. However, these error sources may 
represent systematic errors that behave in a similar way for a
given line in different stars, so that the star-to-star scatter likely
underestimates the real error bars.

While the O abundance is derived from the IR triplet at 7770\AA\, whose lines
are clean (i.e. not blended at this spectral resolution)
and are a well known abundance indicator, the lines are of very
high excitation from the dominant ion.  Thus, as discussed above, 
O/Fe ratios are very sensitive to the adopted temperature and gravity when
O abundances are compared to that given by Fe I lines. O/Fe ratios are much more
robust when Fe II lines are used; unfortunately, our Fe II abundances are
only based on one or two lines.  Oxygen is 
clearly overabundant in NGC 6553 stars, with an abundance ratio similar to that
measured in other globular clusters and metal-poor stars (Gratton \&
Ortolani 1986,  1989; Barbuy 1988). Our O abundance for \cyg\ is also quite high; however, this 
star is known to be a mild Ba star (Sneden, Lambert \& Pilachowski 1981), 
and its surface
abundances may have been modified by accretion of material from an evolved,
originally more massive, companion.

Mg abundances are derived from the line at 8718\AA\ that appears
clean both in the Sun and in all our spectra. Parameters for this line were
taken from Lambert \& Luck (1978) solar analysis,
and give a solar Mg abundance in good agreement with that given by Anders
\& Grevesse (1989).  Mg is then clearly
overabundant in NGC 6553 stars.

Si abundances are obtained from a few (1--4) lines; these are clean
solar lines used in the analysis by Lambert \& Luck (1978). Results given by
different lines are generally internally consistent with each other for each 
star. However, Si abundances are
quite sensitive to the adopted effective temperatures: if \teff\ is
lowered by 100 K (i.e. within uncertainties of the present analysis), [Si/Fe]
increases by 0.12 dex. This sensitivity might explain the rather large
star-to-star scatter. We conclude that our Si abundance for NGC 6553 may have
an error bar as large as $\pm 0.2$ dex.

Our Ca abundances are much more reliable than those for Si, since they are
derived from at least 5 lines in each star giving very consistent results.
Highly accurate laboratory oscillator strengths are used for the Ca lines (Smith
\& Raggett 1981). Also, the sensitivity of
the Ca abundances to atmospheric parameters is quite similar to that of Fe.
Ca is then clearly overabundant in NGC 6553: the average value of 
[Ca/Fe]=$+0.26\pm 0.1$\ is very similar to that obtained by Gratton \& 
Sneden (1991) for both field metal-poor stars
([Ca/Fe]=$+0.29\pm 0.06$) and globular clusters ([Ca/Fe]=$+0.33\pm 0.15$).

Finally, Ti abundances are obtained from only two Ti I lines, with solar
oscillator strengths. However, they give very consistent star-to-star 
results, and the sensitivity to atmospheric parameters of the [Ti/Fe] ratios
is not large (although a little larger than for Ca). On the whole our
[Ti/Fe] ratio should have an error bar of about $\pm 0.15$~dex. Again, our
average [Ti/Fe] ratio for NGC 6553 ([Ti/Fe]=$+0.19\pm 0.15$) agrees well
with the results for both field metal-poor stars ([Ti/Fe]=$+0.28\pm 0.11$) and
globular clusters ([Ti/Fe]=$+0.21\pm 0.16$) obtained by Gratton \& Sneden
(1991).

Summarizing, our analysis consistently gives overabundances for all the
classical $\alpha-$elements in NGC 6553: for those elements having the
most reliable
results (O, Mg, Ti, and Ca), the abundance ratios are very similar
to those obtained in the analysis of field metal-poor stars and other
globular clusters. The error bar for Si is larger, so that we do not
attribute much weight to the lower value of the overabundance we find for this
element. (It still agrees with the value for field stars of 
[Si/Fe]=$+0.30\pm 0.08$\ obtained by Gratton \& Sneden 1991.) Our abundances
agree also quite well with the pattern observed by McWilliam \& Rich (1994)
for field giants in the Baade's Window (although these authors do not find
an O overabundance).

The most common interpretation of the overabundance of O and $\alpha-$elements
in metal-poor stars is that only massive stars exploding as type II SN
contributed initially 
to nucleosynthesis. This seems to be the case also for NGC 6553,
pointing toward a fast formation of at least part of the galactic bulge.

\section{Discussion of Results}

We have carried out a detailed abundance analysis for five RHB stars
in the globular cluster NGC 6553.  We obtain a mean [Fe/H] = 
$-$0.16 dex, with
the total range of the values for the five stars being only 0.23 dex.
The relative abundance for the best determined 
$\alpha$ process element [Ca/Fe] is +0.26 dex.  Because we have
made a major effort to carry out this abundance analysis in a
manner consistent with that adopted by 
Carretta \& Gratton (1997), our results should be directly comparable
with theirs.  Their work represents the largest collection of 
galactic globular
cluster abundances based on a uniform analysis using both their own 
spectra and reanalyzing
\eqw\ from high dispersion analyses in the literature.  We have
thus extended their abundance calibration up to the highest values
of metallicity found among the galactic GCs.

Zinn \& West (1984) assigned [Fe/H] = $-0.29 {\pm}0.11$ dex to NGC 6553.
It is the GC with the highest metallicity among their calibrating
clusters (Table 5 of their paper).  The highest metallicity they
assign to any galactic GC is +0.24 dex to Terzan 1 and to Terzan 5, both heavily
obscured clusters in the galactic bulge.  Their abundance for NGC 6553 is 
remarkably close (0.13 dex smaller) to the results of our abundance analysis.  
It is equal to the systematic difference of +0.12 ${\pm}$0.01 dex found by 
CG between the results of their abundance analysis and those
of previous investigators using the same set of \eqw.

CG compare their uniform well calibrated scale for a limited number
of GCs with that of Zinn \& West (1984), which is
the metallicity scale most commonly used for the family of galactic GCs,
over the full metallicity range spanned by each.  They find a curvature
such that at both low and high metallicities (``high'' here
is M71, 47 Tuc and NGC 6352), the Zinn \& West scale 
appears to be somewhat high (by about 0.1 dex).  The point we have
added for NGC 6553 at the extremely metal rich end
appears to indicate that the comparison of the two metallicity
scales for galactic GCs is somewhat more complex.  

Rutledge \etal\ (1997b) have introduced the parameter
$W(CaII)$ which is determined
from observations of the strength of the infrared Ca triplet in individual
giants from low resolution spectra as the basis for a new metallicity
ranking scheme for the family of galactic globular clusters.
As is shown in Figure 6, with the addition of the point for NGC 6553, the relationship
of the CG abundance scale with this parameter appears to be non-linear. 

A detailed abundance
analysis of a second extremely high metallicity GC seems desirable to confirm 
this.

The average metallicity for the five RHB stars in NGC 6553 is slightly higher than
the average value of approximately $-0.2$ dex found
for giants in Baade's Window in the galactic bulge
by Rich (1988) as recalibrated by McWilliam \& Rich (1994) 
(see also Castro \etal\ 1995) and by Sadler, Rich \& Terndrop (1996).
Castro \etal\ (1996) establish
that the most metal rich star known in Baade's window from this
sample and the most metal rich star known in the
solar neighborhood ($\mu$ Leo) both have [Fe/H] = +0.45 dex, so there may well
be galactic GCs that are even more metal rich than NGC 6553 as suggested
by the Zinn \& West compilation.

\subsection {Comparison with Barbuy \etal\ (1999)}

We have found an abundance for NGC 6553 which is considerably higher
than that of Barbuy \etal\ (1999).
When comparing our abundances with those of
Barbuy \etal, we should recall that there are differences in
procedure which we believe will produce a systematic difference in
the derived abundances.  In particular these lie in the choice of
model atmospheres and in the calculation of the solar
abundances.  Barbuy \etal\ adopt the model atmospheres of Plez \etal\ (1992)
for their two giants in NGC 6553 while we use those of Kurucz (1992).
They adopt the solar model of Holweger \& Muller (1974).
For these reasons, our Fe abundances are
expected to be 0.1 to 0.15 dex higher. This explains part of the 
difference between
our result and that by Barbuy \etal, but not all.  Some perhaps must
be ascribed to our use of hotter stars and the higher SNR and 
dispersion of our spectra, leading to better continuum definition.

\section{Summary}

We provide a high dispersion line-by-line abundance analysis
of five red HB stars in the extremely metal rich
galactic globular cluster NGC 6553.  In such a metal rich cluster,
these red HB stars are significantly hotter
than the red giants near the tip of the giant branch, and hence their
spectra will be much less crowded.  Since accurate  
location of the continuum is the critical key to success for abundance analyses
of such metal rich objects, our approach offers the potential for a 
more reliable abundance determination.

We find that the mean [Fe/H] for NGC 6553 is $-0.16$ dex,
comparable to the mean abundance in the galactic bulge found
McWilliam \& Rich (1994)
and considerably higher than that
obtained from an analysis of two red giants in this cluster by
Barbuy \etal\ (1999).
The relative abundance for the best determined 
$\alpha$ process element  indicates an excess of $\alpha$ 
process elements of a factor of two
as [Ca/Fe] = +0.26 dex.  The metallicity
of NGC 6553 reaches the average
of the Galactic bulge and of the solar neighborhood.  It is likely
that there are even more metal rich galactic globular clusters that 
have solar metallicity.

Our analysis of the abundance of NGC 6553 
provides an important calibration point for
determining the metallicity of the more distant/more heavily reddened
extremely metal rich globular clusters found exclusively in the
nuclear bulge.  It also provides an important clue regarding the 
mean abundance of the stellar population
in the galactic bulge and in luminous elliptical galaxies in general.

\acknowledgements 
We are grateful to Maria Donata Guarnieri and her collaborators for providing
access to their HST and IRAC photometry of NGC 6553 
prior to publication and for providing
finding charts for candidate HB stars from their images and to Jim McCarthy
for help during the 1997 observing run.  We also thank Beatriz Barbuy and Sergio
Ortolani for having provided a copy of their recent paper in advance of
publication. The entire Keck/HIRES user community owes a huge debt
to Jerry Nelson, Gerry Smith, Steve Vogt, and many other people who have
worked to make the Keck Telescope and HIRES a reality and to
operate and maintain the Keck Observatory.  We are grateful
to the W. M. Keck Foundation, and particularly its late president,
Howard Keck, for the vision to fund the construction of the W. M. Keck
Observatory. This research has made use of
the SIMBAD data base, operated at CDS, Strasbourg, France.

\clearpage

%
%
\begin{deluxetable}{lrrrrrrr}
\tablenum{1}
\tablewidth{0pt}
\scriptsize
\tablecaption{The Sample of Stars Observed in NGC 6553}
\label{tab1}
\tablehead{\colhead{ID} & \colhead{$V$\tablenotemark{a}} &  
\colhead{$I$\tablenotemark{a}} &
\colhead{$J$\tablenotemark{a}} & 
\colhead{$K$\tablenotemark{a}} & \colhead{$v_r$} &
\colhead{Date of Obs.} & \colhead{Exp. Time} \nl
\colhead{} & \colhead{(mag)} & \colhead{(mag)} & \colhead{(mag)} & 
\colhead{(mag)} & \colhead{(\kms)} & \colhead{} & \colhead{(sec)}
}
\startdata
Red HB Stars \nl
30180 & 16.85 & 14.86 & 13.39 & 12.46 &  $-$4.9 & 950805 & 3000 \nl
30242 & 16.90 & 14.97 & 13.38 & 12.42 &  +2.2 &  950805 & 3000 \nl
30257 & 16.88 & 14.98 & 13.32 & 12.38 &  +1.6 & 950805 & 3000 \nl
40071 & 16.77 & 14.82 & 13.38 & 12.40 &  $-$4.5  & 970803 & 3000 \nl
40123 & 16.94 & 14.95 & 13.43 & 12.46 & +12.6 & 970803 &  3000 \nl
Ext. Red Giants \nl
30291 & 16.92 & 12.15 & 8.90 & 7.17 &  +9.1 &  950806 &  2000 \nl
30297& 16.62 & 12.04 & 8.81 & 7.31 &  +10.3 &  950806 &  2000 \nl
Comparison Star \nl
\cyg (BS 8115) & 3.19 & & 1.68 & 1.07 & +14.4 & 950805 & 2 \nl
\enddata
\tablenotetext{a}{From Guarnieri \etal\ (1998), except for \cyg, whose
photometry is from Johnson (1964) and with $K$ from Neugebauer \& Leighton (1969).}
\end{deluxetable}

%
%
\begin{deluxetable}{lrrr}
\tablenum{2}
\tablewidth{0pt}
\scriptsize
\tablecaption{Adopted Model Atmosphere Parameters}
\label{tab2}
\tablehead{\colhead{Star ID} & \colhead{\teff} & 
\colhead{log(g)} & \colhead{$v_t$} \nl
\colhead{} & \colhead{(K)} & \colhead{(dex)} &
\colhead{(\kms)}
}
\startdata
Red HB Stars \nl
30180   &  4700 &  2.3 &   1.8 \nl
30242  &   4630 &  2.3  &   1.8 \nl
30257  &   4750\tablenotemark{a} &  2.3  &  1.6 \nl
40071  &  4725 &  2.3  &  2.5 \nl
40123  &  4830\tablenotemark{b}  &  2.3  &  1.4 \nl
Comparison Star \nl
\cyg (BS 8115) &   4950  &  2.7 &   1.6 \nl
\enddata
\tablenotetext{a}{Initial \teff = 4600 K.}
\tablenotetext{b}{Initial \teff = 4625 K.}
\end{deluxetable}

\begin{deluxetable}{lrrrrrrrrrrrrrrrrrr}
\tablenum{3}
\tablewidth{0pt}
\tabcolsep 2pt 
\scriptsize
\tablecaption{Equivalent Widths For 5 Red HB Stars in NGC 6553 and for \cyg\tablenotemark{a}}
\label{tab3}
\tablehead{
\colhead{Ion} & \colhead{~~~~$\lambda$ (\AA)}	&\colhead{$\chi$ (eV)}	&$\log(gf)$
&\mc{400071}	&\mc{40123} 		&\mc{30180}	&\mc{30242} 	&\mc{30257}		&\mc{\cyg} \nl
  & & & & \mc{(m\AA)} & \mc{(m\AA)} & \mc{(m\AA)} & \mc{(m\AA)} & \mc{(m\AA)} & \mc{(m\AA)}  
}
\startdata
O I	&7771.95	&9.11	&0.33		&&56.2		&&	 	&&30.2	 	&&39.7	 	&&56.8		&&49.0		\nl
O I	&7774.18	&9.11	&0.19		&&53.3		&&33.3	 	&&42.7	 	&&44.2	 	&&		&&55.9		\nl
O I	&7775.40	&9.11	&$-0.03$	&&45.9		&&31.7	 	&&39.7	 	&&42.5	 	&&43.3		&&47.8		\nl

Mg I	&8717.80	&5.93	&$-1.09$	&&122.4		&&86.9	 	&&94.9	 	&&101.0	 	&&		&&97.9		\nl

Si I	&6848.57	&5.86	&$-1.75$	&&24.5		&&20.6	 	&&		&&24.1	 	&&34.8	 	&&38.7		\nl
Si I	&7034.90	&5.87	&$-0.88$	&&81.2		&&63.8	 	&&90.3	 	&&		&&60.4	 	&&		\nl
Si I	&7405.79	&5.61	&$-0.82$	&&114.6		&&		&&		&&		&&		&&		\nl
Si I	&7932.40	&5.96	&$-0.47$	&&74.8		&&72.8		&&		&&92.4		&&80.0		&&95.7		\nl

Ca I	&6455.60	&2.52	&$-1.29$	&&		&&108.6	 	&&102.8		&&		&&		&&		\nl
Ca I	&6462.57	&2.52	&0.26		&&334.5		&&230.6	 	&&315.4	 	&&311.1	 	&&311.0	 	&&		\nl
Ca I	&6471.67	&2.52	&$-0.69$	&&173.9		&&133.0	 	&&165.2	 	&&142.3	 	&&161.0	 	&&		\nl
Ca I	&6493.79	&2.52	&$-0.11$	&&251.8		&&	 	&&199.0	 	&&176.9	 	&&170.4	 	&&182.8		\nl
Ca I	&6499.65	&2.52	&$-0.82$	&&158.2		&&116.0	 	&&137.3	 	&&122.9	 	&&134.2	 	&&131.9		\nl
Ca I	&6572.80	&0.00	&$-4.32$	&&186.1		&&108.5	 	&&131.9	 	&&129.7	 	&&137.6	 	&&105.9		\nl

Sc II	&6604.60	&1.36	&$-1.14$	&&91.4		&&74.5	 	&&89.6	 	&&91.3	 	&&68.4	 	&&90.5		\nl

Ti I	&6508.12	&1.43	&$-2.05$	&&41.7		&&41.3		&&42.5	 	&&		&&42.7	 	&&		\nl
Ti I	&6743.13	&0.90	&$-1.63$	&&126.9		&&76.3	 	&&78.5	 	&&112.7		&&		&&69.7		\nl

Cr I	&6979.80	&3.46	&$-0.22$	&&111.2		&&	 	&&90.6		&&		&&		&&		\nl
Cr I	&7400.19	&2.90	&$-0.11$	&&156.3		&&105.1	 	&&147.0	 	&&137.4	 	&&132.8	 	&&131.7		\nl

Fe I	&6380.75	&4.19	&$-1.34$	&&98.2	 	&&80.8		&&91.3	 	&&86.0	 	&&97.4		&&100.8		\nl
Fe I	&6392.54	&2.28	&$-3.97$	&&93.3	 	&&51.3		&&		&&		&&		&&		\nl
Fe I	&6393.61	&2.43	&$-1.43$	&&271.4  	&&167.0  	&&218.1  	&&235.4	 	&&223.7 	&&208.5		\nl
Fe I	&6400.32	&3.60	&$-0.23$	&&345.9 	&&		&&		&&		&&		&&		\nl
Fe I	&6411.66	&3.65	&$-0.60$	&&236.4  	&&125.2  	&&156.8  	&&		&&161.3 	&&168.2		\nl
Fe I	&6421.36	&2.28	&$-1.98$	&&216.4  	&&162.0  	&&222.1 	&&188.2  	&&204.9 	&&		\nl
Fe I	&6481.88	&2.28	&$-2.94$	&&158.7  	&&106.3  	&&142.8  	&&124.4  	&&124.9 	&&132.5		\nl
Fe I	&6498.95	&0.96	&$-4.66$	&&153.2  	&&113.5  	&&143.7  	&&124.5  	&&121.1 	&&134.5		\nl
Fe I	&6518.37	&2.83	&$-2.56$	&&133.6  	&&95.0		&&		&&		&&		&&		\nl
Fe I	&6533.94	&4.56	&$-1.28$	&&92.8		&&		&&		&&		&&		&&		\nl
Fe I	&6574.25	&0.99	&$-4.96$	&&142.6  	&&100.0  	&&122.6 	&&121.9  	&&117.2 	&&		\nl
Fe I	&6581.22	&1.48	&$-4.68$	&&133.0  	&&81.8	 	&&93.6	 	&&108.5  	&&102.5 	&&102.7		\nl
Fe I	&6593.88	&2.43	&$-2.30$	&&183.8  	&&146.0  	&&170.9  	&&170.9  	&&163.7 	&&157.0		\nl
Fe I	&6608.04	&2.28	&$-3.96$	&&92.5	 	&&56.5	 	&&73.8	 	&&55.9	 	&&63.8		&&		\nl
Fe I	&6609.12	&2.56	&$-2.65$	&&161.5  	&&121.4  	&&137.5  	&&136.4  	&&122.9 	&&		\nl
Fe I	&6625.04	&1.01	&$-5.32$	&&		&&		&&		&&		&&114.3 	&&		\nl
Fe I	&6627.56	&4.55	&$-1.50$	&&69.3	 	&&65.2	 	&&	 	&&52.0	 	&&63.7		&&64.7		\nl
Fe I	&6703.58	&2.76	&$-3.00$	&&128.2  	&&88.7	 	&&102.4  	&&92.2	 	&&97.4		&&		\nl
Fe I	&6713.74	&4.79	&$-1.41$	&&		&&32.2	 	&&		&&47.4		&&		&&		\nl
Fe I	&6725.36	&4.10	&$-2.21$	&&43.3	 	&&		&&48.8	 	&&		&&43.9		&&		\nl
Fe I	&6726.67	&4.61	&$-1.05$	&&76.3	 	&&59.0	 	&&63.3	 	&&		&&77.1		&&86.0		\nl
Fe I	&6733.15	&4.64	&$-1.44$	&&52.2	 	&&42.9	 	&&57.0	 	&&46.5	 	&&61.8		&&57.8		\nl
Fe I	&6750.16	&2.42	&$-2.58$	&&		&&		&&155.0  	&&147.7  	&&145.6 	&&128.1		\nl
Fe I	&6786.86	&4.19	&$-1.90$	&&66.0	 	&&		&&64.9		&&		&&		&&		\nl
Fe I	&6839.84	&2.56	&$-3.35$	&&96.6	 	&&83.3	 	&&100.4  	&&		&&102.3 	&&		\nl
Fe I	&6843.66	&4.55	&$-0.86$	&&109.9  	&&84.3	 	&&111.3  	&&74.9	 	&&103.1 	&&98.9		\nl
Fe I	&6857.25	&4.07	&$-2.07$	&&51.6	 	&&36.0	 	&&54.6	 	&&31.1		&&		&&51.7		\nl
Fe I	&6858.16	&4.61	&$-0.95$	&&86.7	 	&&60.2	 	&&76.5	 	&&69.1	 	&&82.5		&&92.3		\nl
Fe I	&6898.29	&4.22	&$-2.08$	&&45.5	 	&&		&&		&&43.7	 	&&		&&48.5		\nl
Fe I	&6916.69	&4.15	&$-1.35$	&&126.9  	&&88.0		&&		&&		&&		&&109.7		\nl
Fe I	&6988.53	&2.40	&$-3.42$	&&126.6  	&&106.0  	&&		&&107.3  	&&115.1 	&&106.2		\nl
Fe I	&7007.97	&4.18	&$-1.80$	&&		&&		&&		&&		&&		&&78.6		\nl
Fe I	&7022.96	&4.19	&$-1.11$	&&121.0  	&&	 	&&	 	&&122.9  	&&120.9 	&&		\nl
Fe I	&7130.93	&4.22	&$-0.76$	&&146.5  	&&142.2 	&&		&&		&&		&&		\nl
Fe I	&7132.99	&4.07	&$-1.66$	&&83.9	 	&&63.6		&&		&&		&&		&&		\nl
Fe I	&7142.52	&4.95	&$-0.93$	&&99.7	 	&&74.3	 	&&		&&79.9	 	&&80.7		&&		\nl
Fe I	&7151.47	&2.48	&$-3.58$	&&103.2  	&&		&&		&&107.5 	&&		&&		\nl
Fe I	&7180.00	&1.48	&$-4.71$	&&147.7  	&&	 	&&96.5	 	&&118.1  	&&111.9 	&&		\nl
Fe I	&7189.16	&3.07	&$-2.77$	&&96.5	 	&&64.0		&&		&&		&&		&&		\nl
Fe I	&7190.13	&3.11	&$-3.28$	&&94.4		&&		&&50.1		&&49.3		&&		&&		\nl
Fe I	&7306.57	&4.18	&$-1.55$	&&63.9	 	&&48.9	 	&&70.2	 	&&63.2	 	&&59.5		&&69.0		\nl
Fe I	&7401.69	&4.19	&$-1.60$	&&94.4		&&		&&		&&		&&		&&		\nl
Fe I	&7411.16	&4.28	&$-0.48$	&&		&&134.9  	&&161.0  	&&143.6 	&&172.8 	&&168.3		\nl
Fe I	&7418.67	&4.14	&$-1.44$	&&121.8  	&&74.3	 	&&91.6	 	&&86.4		&&97.3		&&94.3		\nl
Fe I	&7421.56	&4.64	&$-1.69$	&&30.2	 	&&		&&		&&27.2		&&		&&33.4		\nl
Fe I	&7447.40	&4.95	&$-0.95$	&&64.8		&&		&&		&&		&&		&&		\nl
Fe I	&7461.53	&2.56	&$-3.45$	&&127.4  	&&82.2	 	&&97.4	 	&&90.0		&&82.8		&&		\nl
Fe I	&7491.66	&4.30	&$-1.01$	&&106.0  	&&78.5	 	&&113.4  	&&91.3		&&100.5 	&&82.1		\nl
Fe I	&7568.91	&4.28	&$-0.90$	&&138.8  	&&116.5  	&&		&&		&&137.4 	&&		\nl
Fe I	&7583.80	&3.02	&$-1.93$	&&151.3  	&&124.6  	&&	 	&&169.9  	&&151.5 	&&		\nl
Fe I	&7719.05	&5.03	&$-0.96$	&&71.6	 	&&51.3	 	&&66.7	 	&&		&&57.7		&&54.6		\nl
Fe I	&7723.21	&2.28	&$-3.62$	&&		&&72.5	 	&&123.0  	&&		&&98.8		&&		\nl
Fe I	&7751.11	&4.99	&$-0.74$	&&104.2  	&&71.1	 	&&81.4	 	&&84.2	 	&&		&&87.2		\nl
Fe I	&7807.91	&4.99	&$-0.51$	&&104.5  	&&	 	&&96.1	 	&&94.8	 	&&89.8		&&108.6		\nl
Fe I	&7912.87	&0.86	&$-4.85$	&&168.1  	&&	 	&&	 	&&	 	&&		&&		\nl
Fe I	&7941.10	&3.27	&$-2.29$	&&109.8  	&&86.0	 	&&	 	&&111.7  	&&107.7 	&&82.1		\nl

Fe II	&6416.93	&3.89	&$-2.70$	&&55.0	 	&&45.9	 	&&52.5	 	&&52.7	 	&&50.0		&&53.9		\nl
Fe II	&6456.39	&3.90	&$-2.10$	&&		&&		&&80.8	 	&&		&&		&&		\nl
Fe II	&7449.34	&3.89	&$-3.10$	&&50.9		&&		&&		&&		&&		&&		\nl

Ni I	&6378.26	&4.15	&$-0.82$	&&45.9	 	&&61.5	 	&&63.5	 	&&51.8	 	&&		&&		\nl
Ni I	&6384.67	&4.15	&$-1.00$	&&		&&41.0	 	&&		&&45.9		&&		&&		\nl
Ni I	&6482.81	&1.93	&$-2.78$	&&145.1  	&&95.6	 	&&120.4  	&&121.8 	&&105.0 	&&104.6		\nl
Ni I	&6532.88	&1.93	&$-3.42$	&&94.3	 	&&		&&		&&73.2		&&		&&		\nl 
Ni I	&6586.32	&1.95	&$-2.78$	&&136.4  	&&77.8	 	&&105.4  	&&110.0  	&&93.3		&&97.3		\nl
Ni I	&6635.14	&4.42	&$-0.75$	&&58.4	 	&&49.9	 	&&		&&45.0		&&		&&		\nl
Ni I	&6767.78	&1.83	&$-2.06$	&&161.6  	&&128.1  	&&165.1  	&&142.2  	&&151.9 	&&132.7		\nl
Ni I	&6772.32	&3.66	&$-0.96$	&&81.4	 	&&70.9	 	&&99.4	 	&&77.6	 	&&80.9		&&90.5		\nl
Ni I	&7030.02	&3.54	&$-1.70$	&&63.0	 	&&46.4	 	&&		&&54.1		&&47.6		&&53.9		\nl
Ni I	&7110.91	&1.93	&$-2.91$	&&143.4  	&&102.0  	&&	 	&&113.7 	&&		&&		\nl
Ni I	&7422.29	&3.63	&$-0.29$	&&153.1  	&&127.0  	&&155.2  	&&149.0  	&&147.8 	&&147.3		\nl
Ni I	&7555.61	&3.85	&$-0.12$	&&174.0  	&&125.7  	&&174.9  	&&148.1  	&&142.7 	&&141.9		\nl
Ni I	&7574.05	&3.83	&$-0.61$	&&		&&92.0		&&		&&		&&		&&		\nl
Ni I	&7715.58	&3.70	&$-0.98$	&&86.4		&&		&&		&&		&&		&&		\nl
Ni I	&7727.62	&3.68	&$-0.30$	&&164.7 	&&128.2  	&&143.7  	&&125.7  	&&148.5 	&&133.8		\nl
Ni I	&7797.59	&3.30	&$-0.82$	&&155.4  	&&103.5 	&&141.3  	&&101.3 	&&129.3 	&&112.4		\nl
\enddata
\tablenotetext{a}{As described in the text, all \eqw are measured from convolved spectra.}
\end{deluxetable}

\begin{deluxetable}{llcllcllcllcllcllc}
\tablenum{4}
\tablewidth{0pt}
\tabcolsep 3pt 
\scriptsize
\tablecaption{Abundances for Five Red HB Stars In NGC 6553}
\label{tab4}
\tablehead{\colhead{\bf Ion} && \multicolumn{2}{c}{\bf Star 40071} && \multicolumn{2}{c}{\bf Star 40123} &&
	\multicolumn{2}{c}{\bf Star 30180} && \multicolumn{2}{c}{\bf Star 30242} &&
	\multicolumn{2}{c}{\bf Star 30257} && \colhead{\bf Sun} \nl
\colhead{} && 
\colhead{\# of} & \colhead{abundance} && 
\colhead{\# of} & \colhead{abundance} && 
\colhead{\# of} & \colhead{abundance} && 
\colhead{\# of} & \colhead{abundance} && 
\colhead{\# of} & \colhead{abundance} && \colhead{abund.} \nl
\colhead{} && 
\colhead{lines} & \colhead{(dex)} &&
\colhead{lines} & \colhead{(dex)} && 
\colhead{lines} & \colhead{(dex)} && 
\colhead{lines} & \colhead{(dex)} && 
\colhead{lines} & \colhead{(dex)} && 
\colhead{(dex)} 
}
\startdata
[Fe/H] \nl
Fe I &&50 & $-$0.18 (0.20) &&39 & $-$0.18 (0.15) &&32 & $-$0.14 (0.32) 
&&35 & $-$0.26 (0.22) &&36 & $-$0.03 (0.18) && $-$4.50 \nl
Fe II &&2 & $-$0.11 (0.22) && 1 & $-$0.33 && 2 & $-$0.18 (0.03) && 1 & $-$0.07 
&&1 & $-$0.20 && $-$4.56 \nl
 & \nl
[A/Fe] \nl
O I\tablenotemark{a}
&&3	& +0.53 (0.07)	&&2	& +0.34 (0.13)	&&3	& +0.41 (0.32)	
&&3	& +0.56 (0.22)	&&2	& +0.68 (0.03)	&& +1.18 \nl
Mg I	&&1	& +0.50		&&1	& +0.34		&&1	& +0.31		
&&1	& +0.48 		&&	&		&& +0.04 \nl
Si I	&&4	& +0.07 (0.22)	&&3	& $-$0.01 (0.13)	&&1	& +0.39		
&&2	& +0.27 (0.05)	&&3	& $-$0.03 (0.24)	&& +0.04 \nl
Ca I	&&5	& +0.36 (0.26)	&&5	& +0.33 (0.12)	&&6	& +0.23 (0.19)	
&&5	& +0.13 (0.12)	&&5	& +0.26 (0.24)	&& $-$1.34 \nl
Sc II\tablenotemark{a}	&&1	& $-$0.16	&&1	& $-$0.13	&&1	& $-$0.04
&&1	& +0.11 		&&1	& $-$0.38		&& $-$4.48 \nl
Ti I	&&2	& +0.18 (0.12)	&&2	& +0.23 (0.3)7	&&1	& +0.24		
&&1	& +0.10		&&1	& +0.19		&& $-$2.52 \nl
Cr I	&&2	& +0.10 (0.22)	&&1	& $-$0.09	&&2	& +0.14 (0.01)	
&&1	& +0.04		&&1	& +0.01		&& $-$1.88 \nl
Ni I	&&14	& $-$0.04 (0.23)  &&14	& +0.02 (0.16)	&&9	& +0.11 (0.17)	
&&14	& +0.02 (0.21)	&&9	& $-$0.06 (0.17) && $-$1.25 \nl
\enddata
\tablenotetext{a}{O I and Sc II are calculated with respect to Fe II, all other ions are with
respect to Fe I.}
\end{deluxetable}

%
%
\begin{deluxetable}{lclrr}
\tablenum{5}
\tablewidth{0pt}
\scriptsize
\tablecaption{Mean Abundances for NGC 6553 and Comparison with Baade's Window Results}
\label{tab5}
\tablehead{
Ion & \colhead{NGC 6553 Mean} & \colhead{$\sigma$} 
& \colhead{Mean BW \tablenotemark{a}} &
\colhead{ {$\sigma$}\tablenotemark{a}} \nl
 & \colhead{(dex)} & \colhead{(dex)} & \colhead{(dex)} & \colhead{(dex)} 
}
\startdata
[Fe/H] \nl
Fe I & $-$0.16 & 0.08 & $-$0.33 \nl
Fe II & $-$0.18 & 0.10 \nl
 & \nl
[A/Fe] & \nl
O I & +0.50 & 0.13 & +0.03 & 0.18 \nl
Mg I & +0.41 & 0.10 & +0.35 & 0.14 \nl
Si I & +0.14 & 0.18 & +0.18 & 0.24 \nl
Ca I & +0.26 & 0.09 & +0.14 & 0.17 \nl
Sc II & $-$0.12 & 0.18 & +0.29 & 0.20 \nl
Ti I & +0.19 & 0.06 & +0.34 & 0.10 \nl
Cr I & +0.04 & 0.09 & $-$0.04 & 0.19 \nl
Ni I & +0.01 & 0.07 & $-$0.04 & 0.08 \nl
\enddata
\tablenotetext{a}{Abundances for Baade's Window are from the 11 giants
studied by McWilliam \& Rich (1994).}
\end{deluxetable}

%
%
\begin{deluxetable}{lrlrlr}
\tablenum{6}
\tablewidth{0pt}
\scriptsize
\tablecaption{Abundances for \cyg}
\label{tab6}
\tablehead{\colhead{Ion} & \colhead{\# of} & \colhead{Original Spectrum} & 
\colhead{\# of} & \colhead{Convolved Spectrum} & \colhead{Solar Abund.} \nl
\colhead{} & \colhead{Lines} & \colhead{Abundance} & 
\colhead{Lines} & \colhead{Abundance}  \nl
\colhead{} & \colhead{} & \colhead{(dex)} & \colhead{} & 
\colhead{(dex)} &  \colhead{(dex)} 
}
\startdata
[Fe/H]  \nl
 Fe I & 43 & +0.08 (0.12) & 27 & +0.05 (0.20) & $-$4.50 \cr
 Fe II & 3 & +0.09 (0.17) & 1 & $-$0.06 & $-$4.56 \cr
 & \nl
[A/Fe] \nl
   O I   &  4 & +0.35 (0.14)   &  3 & +0.38 (0.18)   & +1.18 \nl
   Mg I  &  1 & +0.15          &  1 & +0.20          & +0.04 \nl
   Si I  &  3 & +0.10 (0.11)   &  2 & +0.11 (0.26)   & +0.04 \nl
   Ca I  &  5 & +0.08 (0.14)   &  3 & +0.07 (0.04)   & $-$1.34 \nl
   Sc II &  1 & +0.30          &  1 & +0.12          & $-$4.48 \nl
   Ti I  &  2 & $-$0.08 (0.16) &  1 & $-$0.28        & $-$2.52 \nl
   Cr I  &  4 & +0.03 (0.06)   &  1 & +0.02          & $-$1.88 \nl
   Ni I  & 16 & +0.02 (0.14)   &  9 & $-$0.08 (0.17) & $-$1.25 \nl
\enddata
\end{deluxetable}

\clearpage

\begin{figure}
\epsscale{0.9}
\plotone{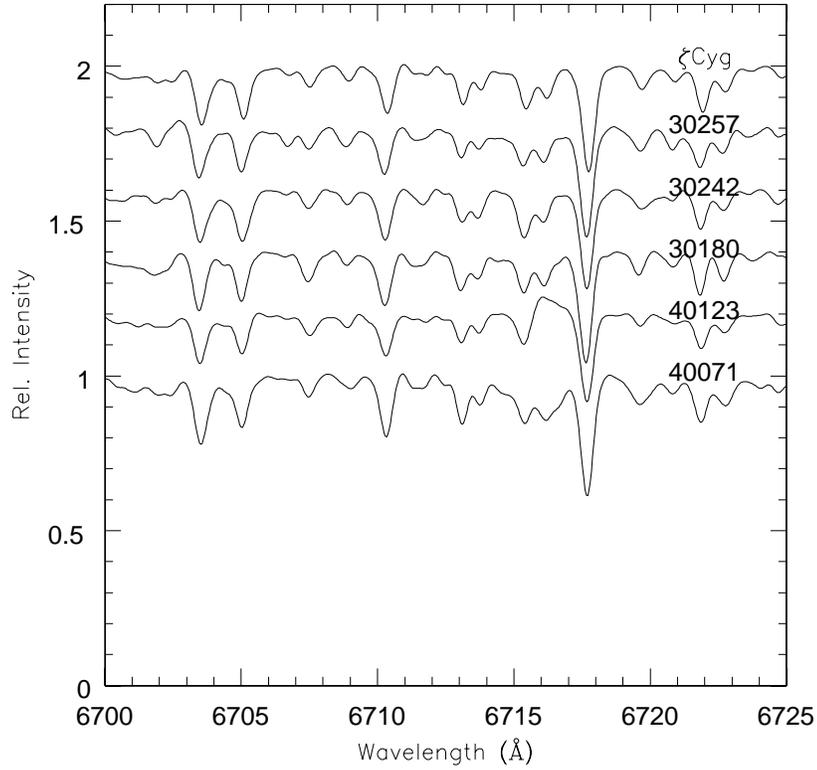}
\caption[figure1.ps]{A portion of the (convolved) spectra is shown for the five
NGC 6553 RHB stars and for the comparison star \cyg.  
The spectra are arbitrarily shifted with respect
to that of star 40071 along the Y axis to avoid confusion.
\label{fig1}}
\end{figure}

\begin{figure}
\epsscale{1.0}
\plotone{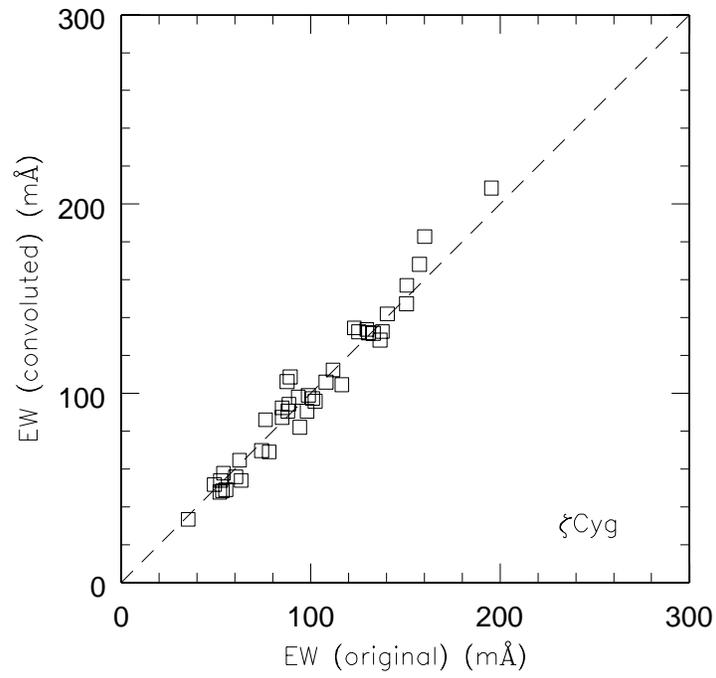}
\caption[figure2.ps]{The set of \eqw\ measured on the original \cyg\ spectrum are plotted
against those measured on the convolved spectrum.  The dashed line corresponds
to equality.
\label{fig2}}
\end{figure}

\begin{figure}
\epsscale{0.9}
\plotone{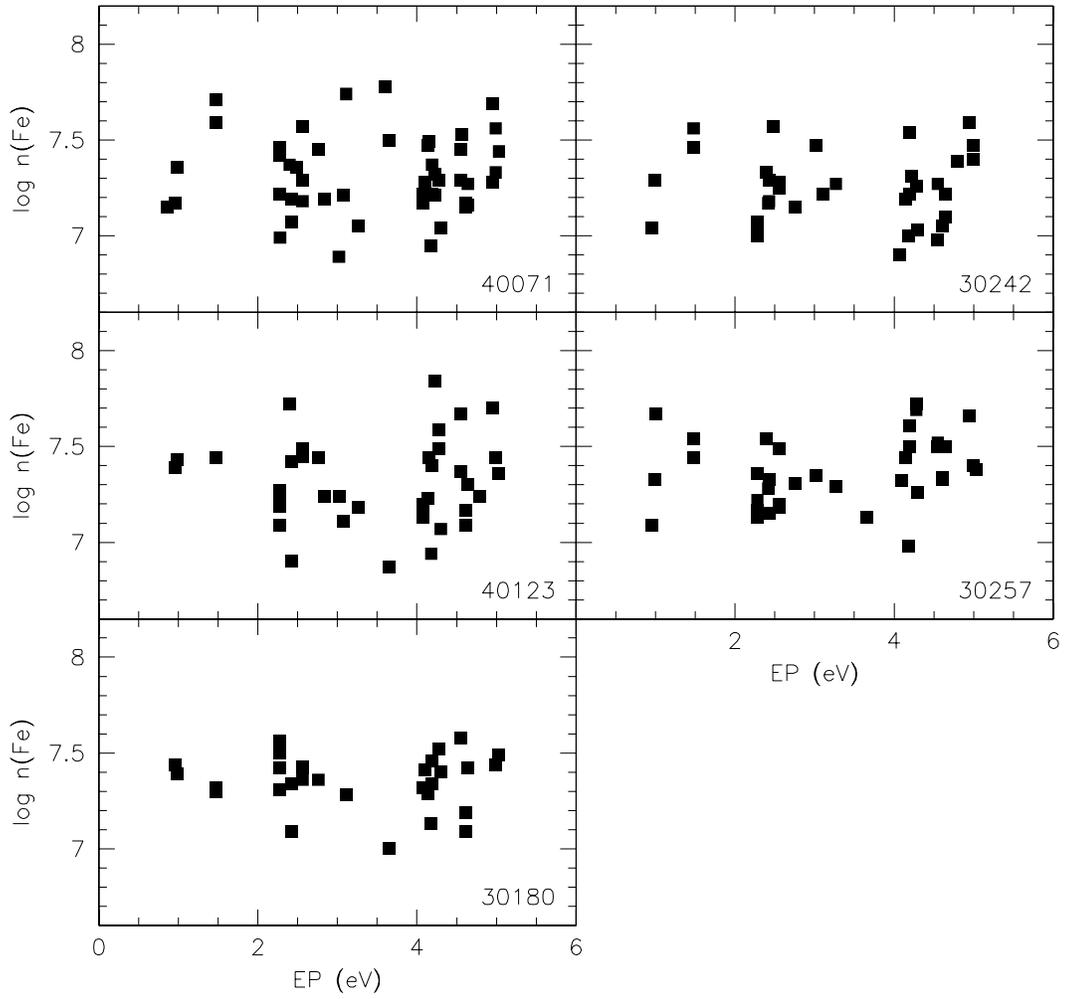}
\caption[figure3.ps]{The abundances deduced from lines of Fe I are shown 
as a function of excitation potential of the lower level of the transition
for the five 
NGC 6553 RHB stars.
\label{fig3}}
\end{figure}

\begin{figure}
\epsscale{0.9}
\plotone{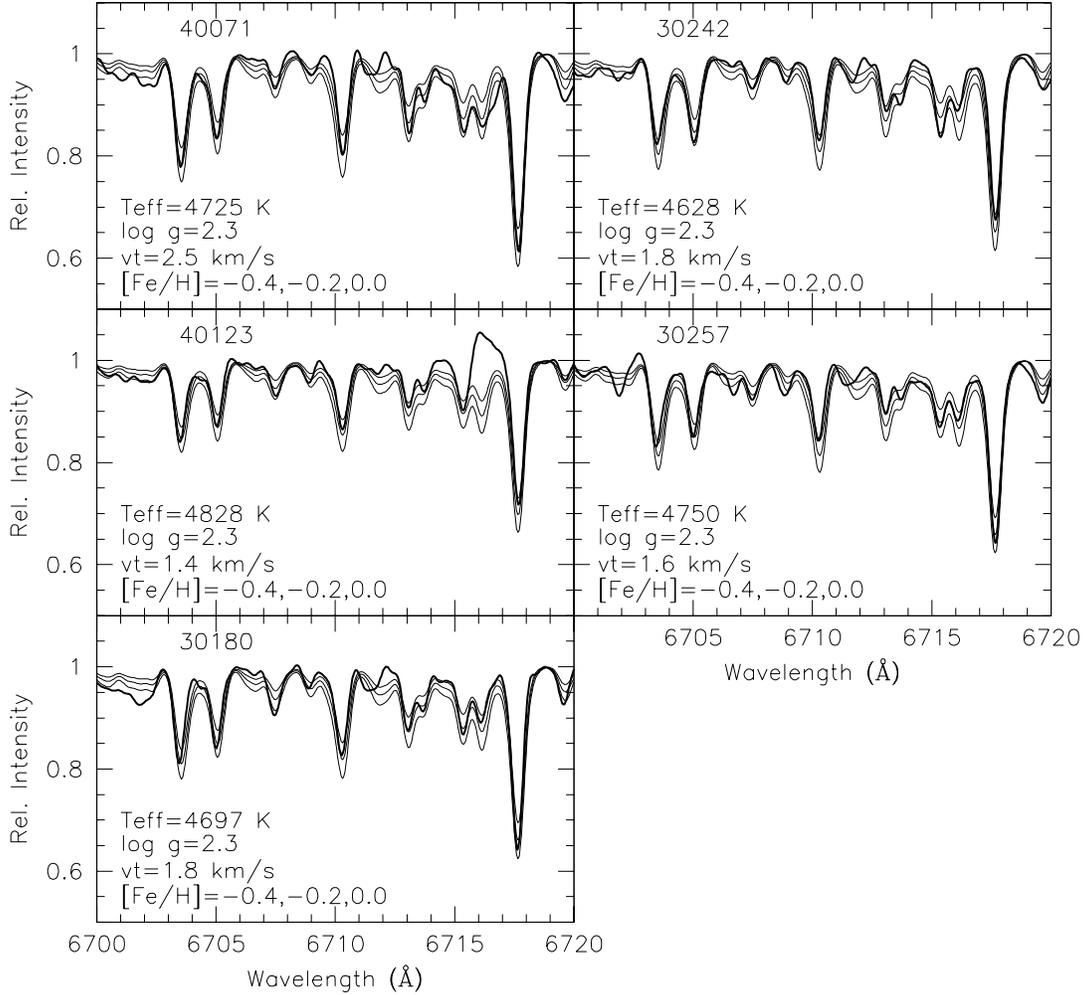}
\caption[figure4.ps]{A spectral synthesis of the region shown in Figure 1 for the
five RHB stars in NGC 6553.  The convolved observed spectra are shown
by the thick lines, while the thin lines denote predictions for model
atmospheres with the parameters indicated in the lower left corner of
each box and for abundances of [Fe/H] = $-0.4$, $-0.2$, and 0.0 dex (i.e. 
solar abundance).
\label{fig4}}
\end{figure}

\begin{figure}
\epsscale{0.9}
\plotone{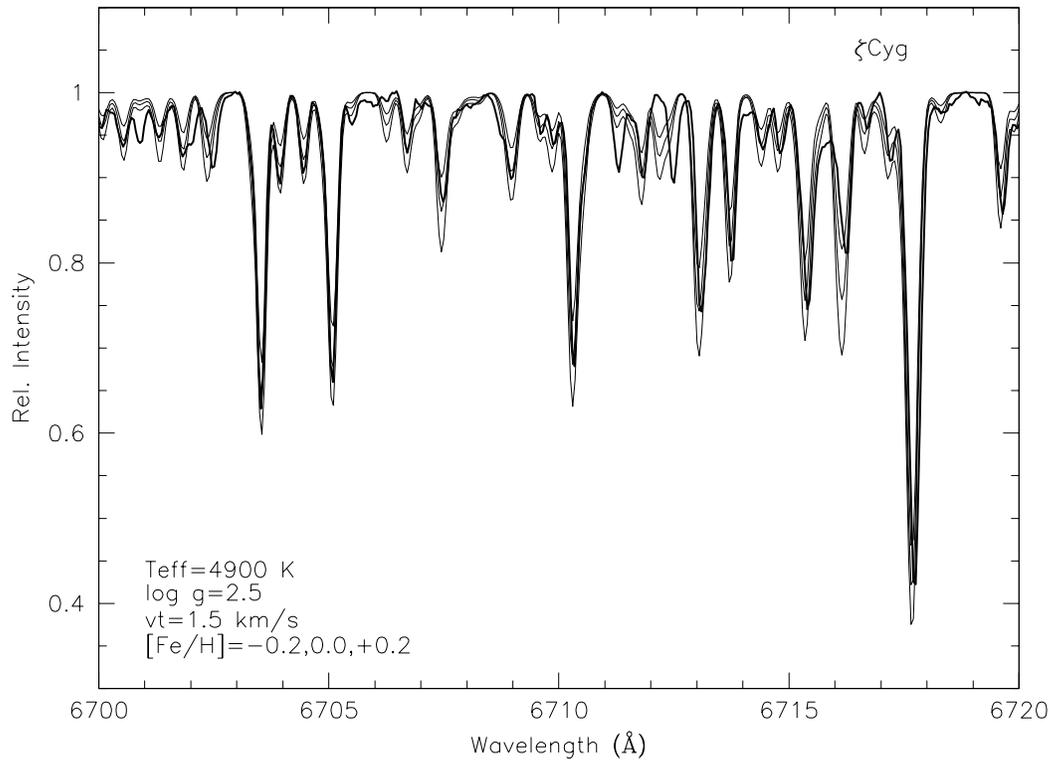}
\caption[figure5.ps]{The same as Figure 4 for \cyg.  The original observed
spectrum is shown prior to convolution.  Here the abundances for the
synthetic spectra are [Fe/H] $-0.2$, 0.0 and +0.2 dex.
\label{fig5}}
\end{figure}

\begin{figure}
\epsscale{0.9}
\plotone{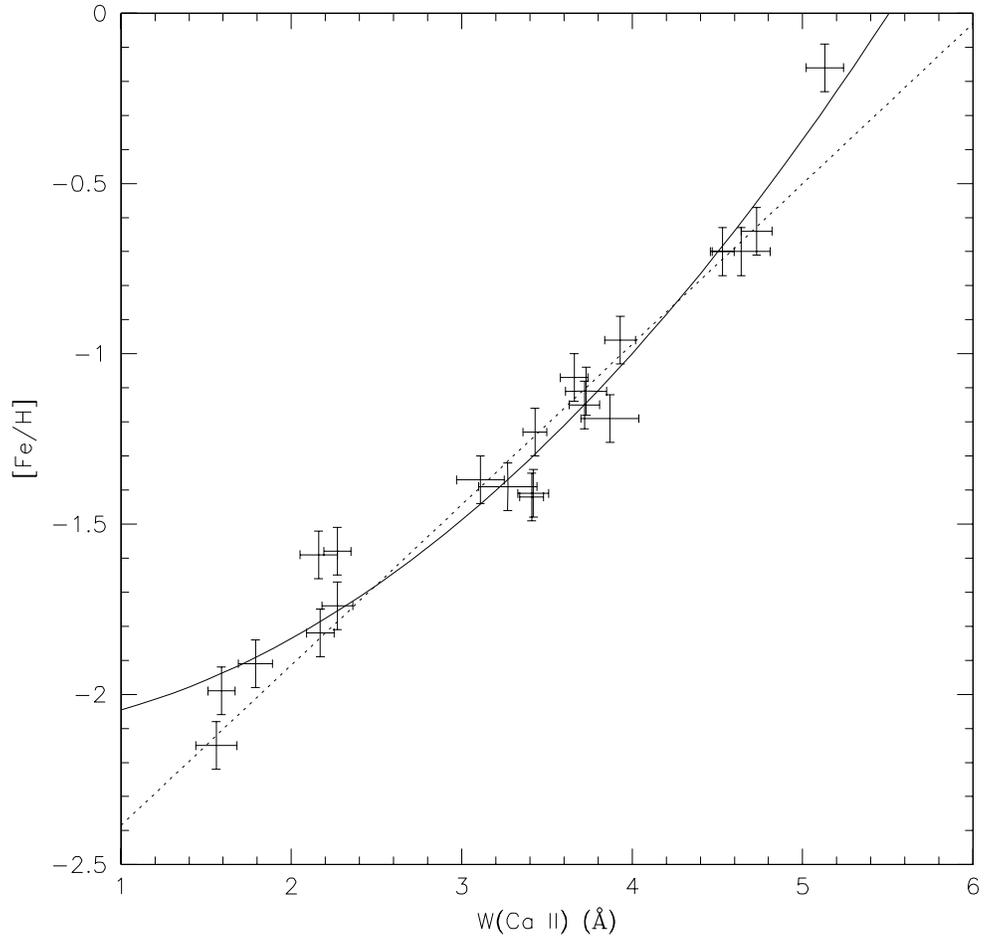}
\caption[figure6.ps]{The parameter $W(CaII)$ defined by Rutledge \etal\ (1997b)
is shown as a function of abundance on the scale of Carreta \& Gratton (1997)
for galactic GCs with high dispersion analyses.  With the addition of
the point representing NGC 6553,
a quadratic fit appears to be superior to a linear fit.
\label{fig6}}
\end{figure}

\end{document}